\documentclass{PoS}
\usepackage{amsmath,amssymb}
\usepackage{bm}
\usepackage{ifthen}
\usepackage{url}
\title{%
\bm{$B$}-meson semileptonic form factors on (2+1+1)-flavor HISQ ensembles
}
\ShortTitle{%
$B$-meson semileptonic form factors on (2+1+1)-flavor HISQ ensembles
}
\author{%
\speaker{Z.~Gelzer}$^{a}$\thanks{Email: zgelzer@illinois.edu},
C.~DeTar$^{b}$,
A.X.~El-Khadra$^{a,c}$,
E.~G\'amiz$^{d}$,
Steven Gottlieb$^{e}$,
Andreas S.~Kronfeld$^{c,f}$,
Yuzhi Liu$^{e}$,
Y.~Meurice$^{g}$,
J.N.~Simone$^{c}$,
D.~Toussaint$^{h}$, and
R.S.~Van de Water$^{c}$
\\
\llap{$^a$}%
Department of Physics,
University of Illinois,
Urbana, IL 61801, USA \\
\llap{$^b$}%
Department of Physics and Astronomy,
University of Utah,
Salt Lake City, Utah 84112, USA \\
\llap{$^c$}%
Fermi National Accelerator Laboratory,
Batavia, Illinois, USA \\
\llap{$^d$}%
CAFPE and Departamento de F\'isica Te\'orica y del Cosmos,
Universidad de Granada,
E-18071 Granada, Spain \\
\llap{$^e$}%
Department of Physics,
Indiana University,
Bloomington, Indiana 47405, USA \\
\llap{$^f$}%
Institute for Advanced Study,
Technische Universit\"at M\"unchen,
D-85748 Garching, Germany \\
\llap{$^g$}%
Department of Physics and Astronomy,
University of Iowa,
Iowa City, IA 52242, USA \\
\llap{$^h$}%
Department of Physics,
University of Arizona, Tucson,
Arizona 85721, USA \\
}
\author{(Fermilab Lattice and MILC Collaborations)}
\abstract{%
We report updates to an ongoing lattice-QCD calculation of the form factors for
the semileptonic decays
$B   \to \pi \ell \nu$,
$B_s \to K   \ell \nu$,
$B   \to \pi \ell^+ \ell^-$, and
$B   \to K   \ell^+ \ell^-$.
The tree-level decays $B_{(s)} \to \pi (K) \ell \nu$
enable precise determinations of the CKM matrix element $|V_{ub}|$, while
the flavor-changing neutral-current interactions $B \to \pi (K) \ell^+ \ell^-$
are sensitive to contributions from new physics.
This work uses MILC's (2+1+1)-flavor HISQ ensembles
at approximate lattice spacings between $0.057$ and $0.15$~fm,
with physical sea-quark masses on four out of the seven ensembles.
The valence sector is comprised of a clover $b$~quark (in the Fermilab
interpretation) and HISQ light and $s$~quarks.
We present preliminary results for the form factors $f_0$, $f_+$, and $f_T$,
including studies of systematic errors.
}
\FullConference{The 37th Annual International Symposium on Lattice Field Theory
                (Lattice 2019)\\
                16--22 June 2019\\
                Central China Normal University, Wuhan, China.}
\newcommand{\para}{\parallel}
\newcommand{\SU}[1]{\mathrm{SU(#1)}}
\newcommand{\NLO}[1]{\ifthenelse{#1 > 1}{\mathrm{N}^{#1}\mathrm{LO}}{\mathrm{NLO}}}
\DeclareMathAlphabet{\mathcal}{OMS}{cmsy}{m}{n}
\begin{document}

\section{Introduction}\label{sec:intro}

These proceedings update an ongoing lattice-QCD calculation \cite{Gelzer:2017edb}
of form factors for the semileptonic decays of $B$ and $B_s$~mesons to pions and
kaons.
Continuing from Ref.~\cite{Gelzer:2017edb}, we calculate the form factors
$f_0$, $f_+$, and $f_T$ for the decays
$B   \to \pi \ell \nu$,
$B_s \to K   \ell \nu$,
$B   \to \pi \ell^+ \ell^-$, and
$B   \to K   \ell^+ \ell^-$.
A major goal of this work is to improve the Standard-Model determination of the
CKM matrix element $|V_{ub}|$, as informed by the tree-level charged-current
decays $B_{(s)} \to \pi (K) \ell \nu$.
The LHCb experiment is expected to publish results for its ongoing study of
$B_s \to K \ell \nu$ \cite{Ciezarek:2016lqu}, while the Belle II experiment
\cite{Urquijo:2015qsa} is also expected to study these decays.

\begin{table}[b]
    \centering
    \caption{%
        Simulation details of the (2+1+1)-flavor MILC ensembles used in this work.
        Columns are (from left to right):
            approximate lattice spacing,
            lattice size,
            tuned quark masses in lattice units,
            tuned $b$-quark hopping parameter, and
            number of configurations.
    }
    \label{tab:sim}
    \begin{tabular}{*{7}{c}}
    \hline\hline
        $\approx a ~(\mathrm{fm})$ &
        $N_s^3 \times N_t$ &
        $a m_l^\prime$ &
        $a m_s^\prime$ &
        $a m_c^\prime$ &
        $\kappa_b^\prime$ &
        $N_\mathrm{cfg}$ \\
    \hline
        $0.15$ &
        $32^3 \times 48$ &
        $0.002426$ &
        $0.06730$ &
        $0.8447$ &
        $0.07732$ &
        $3630$ \\
    \hline
        $0.12$ &
        $24^3 \times 64$ &
        $0.0102$ &
        $0.0509$ &
        $0.635$ &
        $0.08574$ &
        $1053$ \\
        ~ &
        $32^3 \times 64$ &
        $0.00507$ &
        $0.0507$ &
        $0.628$ &
        $0.08574$ &
        $1000$ \\
        ~ &
        $48^3 \times 64$ &
        $0.001907$ &
        $0.05252$ &
        $0.6382$ &
        $0.08574$ &
        $986$ \\
    \hline
        $0.088$ &
        $48^3 \times 96$ &
        $0.00363$ &
        $0.0363$ &
        $0.430$ &
        $0.09569$ &
        $1017$ \\
        ~ &
        $64^3 \times 96$ &
        $0.0012$ &
        $0.0363$ &
        $0.432$ &
        $0.09569$ &
        $1535$ \\
    \hline
        $0.057$ &
        $96^3 \times 192$ &
        $0.0008$ &
        $0.022$ &
        $0.260$ &
        $0.10604$ &
        $1027$ \\
    \hline\hline
    \end{tabular}
\end{table}

This work uses ensembles generated by the MILC collaboration \cite{Bazavov:2012xda},
with (2+1+1) flavors of HISQ sea quarks and one-loop improved L\"uscher--Weisz
gluons.
For the valence sector, we use HISQ light and strange quarks, whose masses match
those of corresponding quarks in the sea, along with clover bottom quarks in the
Fermilab interpretation.
Basic properties of the ensembles are listed in Table~\ref{tab:sim}.

\section{Chiral-continuum fits}\label{sec:chifits}

The form factors on each ensemble and for each recoil energy are obtained from a
correlation-function fitting procedure that is explained in Ref.~\cite{Gelzer:2017edb},
to which we refer the reader.
In comparison to these earlier results, we have:
\begin{itemize}
    \item increased the statistics on the $0.088$-fm ensemble with physical-mass quarks,
    \item added a new $0.088$-fm ensemble with heavier-than-physical light quarks, and
    \item added a new $0.057$-fm ensemble with physical-mass quarks.
\end{itemize}

We match the lattice currents $J$ to those in the continuum $\mathcal{J}$
using a mostly nonperturbative renormalization procedure $\mathcal{J} \doteq Z_J J$,
where $Z_{J_{bq}} = \rho_{J_{bq}} \sqrt{Z_{V^4_{bb}} Z_{V^4_{qq}}}$
\cite{Lepage:1992xa, ElKhadra:2001rv}.
The nonperturbative flavor-diagonal $Z_{V^4}$ factors are provided in
Table~\ref{tab:ZV4}.
The perturbative $\rho_J$ factors are being calculated to one-loop order by
collaborators and will be provided in a blinded form when they are available.

\begin{table}
    \centering
    \caption{%
        Nonperturbative flavor-diagonal current-renormalization factors.
    }
    \label{tab:ZV4}
    \begin{tabular}{*{3}{c}}
    \hline\hline
        $\approx a ~(\mathrm{fm})$ &
        $Z_{V^4_{ss}}$ &
        $Z_{V^4_{bb}}$ \\
    \hline
        $0.15$ &
        $1.9785(31)$ & 
        $0.57617(60)$ \\
    \hline
        $0.12$ &
        $1.9915(18)$ & 
        $0.52492(67)$ \\
    \hline
        $0.088$ &
        $1.9851(29)$ & 
        $0.46560(56)$ \\
    \hline
        $0.057$ &
        $1.9951(42)$ & 
        $0.40136(82)$ \\
    \hline\hline
    \end{tabular}
\end{table}

The form factors are extrapolated to the continuum and corrected for slight
mistunings of the sea-quark masses in a combined chiral-continuum fit, using
heavy-meson rooted-staggered chiral perturbation theory (HMrS$\chi$PT)
\cite{Aubin:2007mc, Bijnens:2010ws, Bazavov:2014wgs}.
We follow the procedure of Ref.~\cite{Bazavov:2019aom}, including both $\SU{2}$
and $\SU{3}$ formulae with terms up to next-to-next-to-next-to-leading order
($\NLO{3}$).

\begin{figure}[t]
    \centering
    \includegraphics[width=\textwidth]{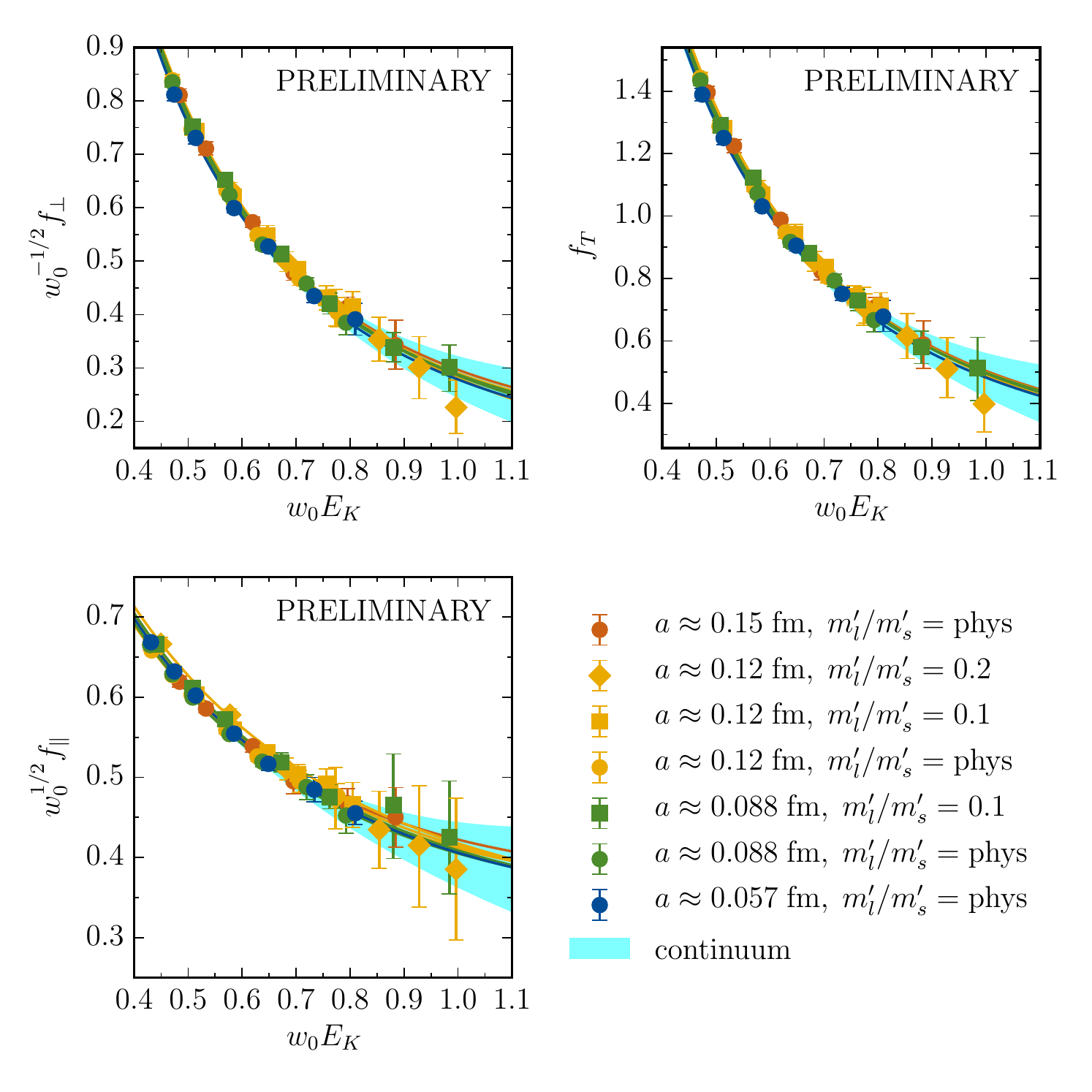}
    \caption{%
        Preliminary chiral-continuum fits for the blinded form factors
        $f_\perp$, $f_T$, and $f_\para$ of the decay $B_s \to K$ as a function
        of the recoil energy $w_0 E_K$, using $\protect\NLO{2}$ $\SU{2}$
        HMrS$\chi$PT.
        Colors denote the lattice spacings and symbols denote the ratios of
        sea-quark masses.
        Colored lines show the fit results evaluated at the parameters of the
        corresponding ensembles.
        The cyan band shows the fit results in the chiral continuum.
    }
    \label{fig:chifit}
\end{figure}

As an example of our central fits, we now focus on the form factors from
$\NLO{2}$ $\SU{2}$ HMrS$\chi$PT, whose preliminary results for $B_s\to K$ are
shown in Fig.~\ref{fig:chifit}.
The expressions are as follows:
\begin{align}
    W_J f_J &= f_J^{(0)} \times \left( c_J^0 \left[ 1 + \delta\!f_J^\mathrm{logs} \right] +
        \delta\!f_J^\mathrm{NLO} + \delta\!f_J^\mathrm{N^2LO} + \cdots \right)
        \times \left( 1 + \delta\!f_J^b \right) ,
    \label{eq:chiff} \\
    f_J^{(0)} &= \frac{g_\pi}{w_0^2 f_\pi (E_L + \Delta_{B^\ast\!})} ,
    \label{eq:chiff0} \\
    \delta\!f_J^\mathrm{NLO} &= c_J^l \chi_l + c_J^s \chi_s + c_J^E \chi_E +
        c_J^{E^2\!} \chi_E^2 + c_J^{a^2\!} \chi_{a^2\!} ,
    \label{eq:chilog} \\
    \delta\!f_J^\mathrm{N^2LO} &= \sum_{m,\,n \in \{l,\,s,\,E,\,E^2,\,a^2\}}
        c_J^{mn} \chi_m \chi_n ,
    \label{eq:chilog2}
\end{align}
where the prefactor $W_J = \{ w_0^{-1/2}, w_0^{1/2}, 1 \}$ for
$J = \{ \perp, \para, T \}$ accounts for the lattice units using the
gradient-flow quantity $w_0$ \cite{Borsanyi:2012zs}, whose continuum value is
$w_0 = 0.1714(15)$~fm \cite{Bazavov:2015yea}.
In Eq.~(\ref{eq:chiff}), $\delta\!f_J^b$ accounts for $b$-quark discretization
effects \cite{Bazavov:2011aa}, while $\delta\!f_J^\mathrm{logs}$ denotes
nonanalytic functions of the lattice spacing and light-quark masses.
$E_L$, along with its abbreviation $E$, is the recoil energy for $L = \{\pi, K\}$.
In Eq.~(\ref{eq:chiff0}), the pole term $\Delta_{B^\ast}$ arises from low-lying
excited states $B^\ast_{(s)(0)}$ as follows:
\begin{align}
    \Delta_{B^\ast} (B_{(s)} \to L) \equiv \frac{M^2_{B^\ast} - M^2_{B_{(s)}} - M^2_L}{2 M_{B_{(s)}}} .
    \label{eq:poles}
\end{align} 
The $B^\ast$ states relevant to $f_\perp$ and $f_T$ have $J^P=1^-$, while those
relevant to $f_0$ have $J^P=0^+$.
For $B_s \to K$, these states are the $B^\ast$ and $B^\ast_0$~mesons, respectively.
The $B^\ast_{0}$~meson has not yet been observed experimentally and is thus
taken from a theoretical determination \cite{Gregory:2010gm}.

Lastly, we incorporate the perturbative current-renormalization factors $\rho_J$
into the HMrS$\chi$PT formulae with the following priors:
\begin{align}
    \tilde{\rho}_J = 1 + \tilde{\rho}_J^{(1)} \alpha_s + \tilde{\rho}_J^{(2)} \alpha_s^2 ,
    \label{eq:rho}
\end{align}
where $\tilde{\rho}_J^{(i)} = 0(1)$.
These priors were chosen to encompass the one-loop calculations for the
corresponding asqtad currents \cite{Bazavov:2019aom, Bailey:2015dka, Bailey:2008wp}
in the absence of the ongoing HISQ calculations for $\rho_J^{(1)}$.

\section{Error budget}\label{sec:errors}

\begin{figure}[t]
    \centering
    \includegraphics[width=0.67\textwidth]{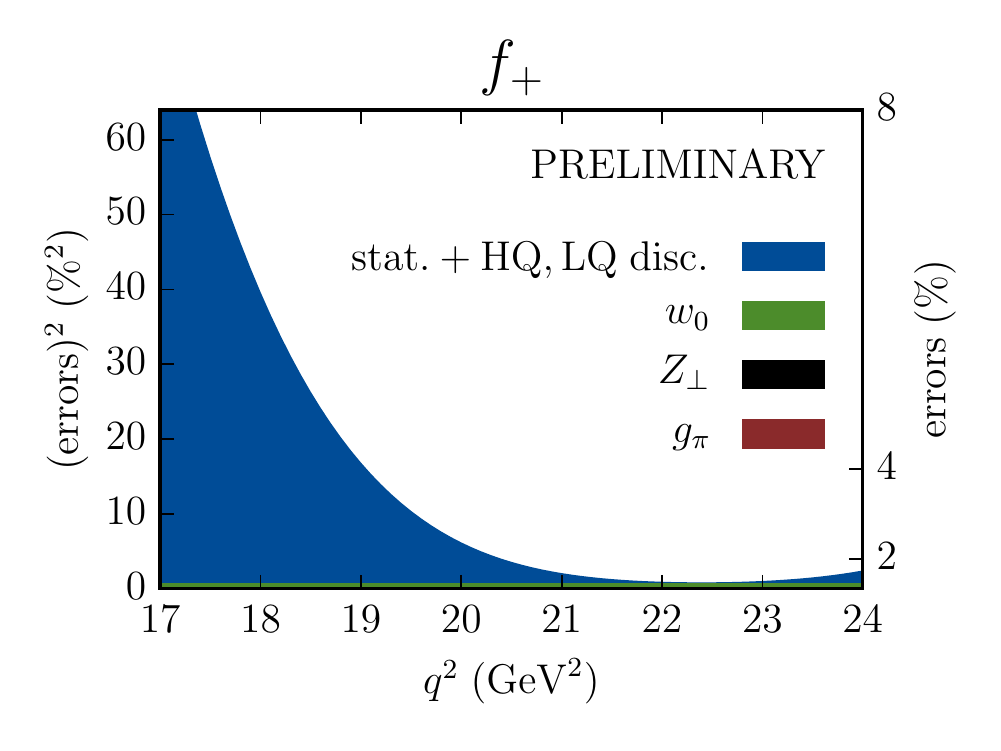}
    \caption{%
        Preliminary distribution of errors for the blinded form factor $f_+$ of
        the decay $B_s \to K$ as a function of the momentum transfer $q^2$.
        The squares of the errors added in quadrature are shown on the left y
        axis, while the errors themselves are shown on the right y axis.
        The blue band shows errors due to statistics, $\chi$PT, and
        light-, strange-, and heavy-quark discretizations, in which the dominant
        contribution is that of statistics.
        The other bands contribute at the subpercent level.
    }
    \label{fig:err}
\end{figure}

We are studying sources of error in order to construct a complete error budget
over the range in $q^2$ for which we have lattice simulations,
$17~(\mathrm{GeV})^2 \lesssim q^2 \lesssim 24~(\mathrm{GeV})^2$.
The form factors from the chiral-continuum fits, as explained in
Sec.~\ref{sec:chifits}, include errors due to to statistics, $\chi$PT, light-,
strange-, and bottom-quark discretizations, current-renormalization factors $Z_J$,
the scale-setting via $w_0$, and the $B^\ast B \pi$ coupling constant $g_\pi$.
The distribution of these errors is shown in Fig.~\ref{fig:err}, where the
dominant source is that of statistics.
The uncertainties due to current renormalization include the contributions from
the fit posteriors of the perturbative factors in Eq.~(\ref{eq:rho}).

\begin{figure}[t]
    \centering
    \includegraphics[width=0.67\textwidth]{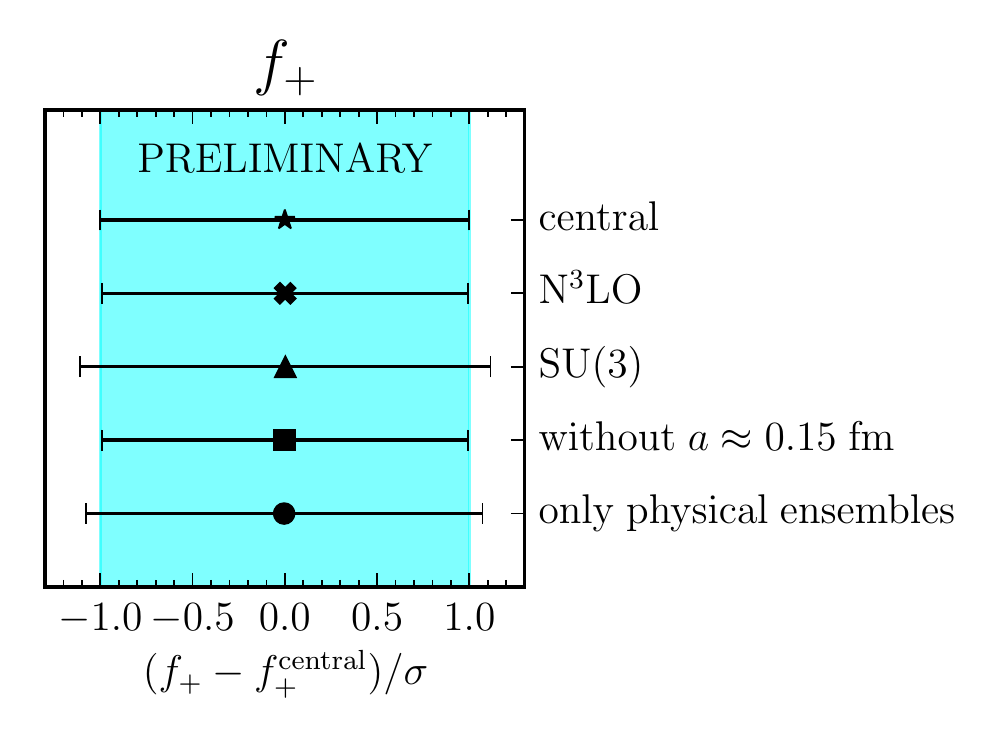}
    \caption{%
        Preliminary chiral-continuum fit variations for the blinded form factor
        $f_+$ of the decay $B_s \to K$ compared with the central fit, whose
        standard error is given by $\sigma$.
        The comparisons are evaluated at a recoil energy of $E_K \approx 900$~MeV,
        which corresponds to $w_0 E_K \approx 0.78$ or
        $q^2 \approx 19.4 ~(\mathrm{GeV})^2$.
        From top to bottom, the HMrS$\chi$PT fit variations are as follows:
        the star denotes the central fit, $\protect\NLO{2}$ $\SU{2}$;
        the x denotes $\protect\NLO{3}$ $\SU{2}$;
        the triangle denotes $\protect\NLO{2}$ $\SU{3}$;
        the square denotes excluding the coarsest ensemble; and
        the circle denotes including only the physical-mass ensembles.
    }
    \label{fig:stab}
\end{figure}

To estimate the truncation effects of our chiral-continuum fits, we consider
variations in both the fit formulae and the lattice data.
Such variations are exemplified in Fig.~\ref{fig:stab}, which compares the
form-factor results at a recoil energy of $E_K \approx 900$~MeV.
This probes the higher momenta on the lattice while remaining within the region
of validity of $\chi$PT.

Including terms up to $\NLO{3}$ in Eq.~(\ref{eq:chiff}) provides results that
agree with those of $\NLO{2}$, indicating that higher-order truncation effects
are negligible.
Next we consider using $\SU{3}$ formulae, which include the effects of
dynamical strange quarks on the nonanalytic functions in Eq.~(\ref{eq:chiff}).
We find $\SU{3}$ to be consistent with $\SU{2}$.

We also consider the effects of reducing the lattice data used in the fits.
Excluding the coarsest ensemble (at $a \approx 0.15$~fm) typically has no
significant effect on the fit results.
Including only the four physical-mass ensembles has small but different effects
for each form factor, which warrants further investigation.

We are actively studying other sources of error.
The simulations in this work used well-tuned parameters for quark masses,
including the bottom-quark hopping parameter.
Preliminary estimates of the corrections due to these uncertainties are all
below $0.1\%$.
We also plan to quantify the finite-volume effects by comparing to the
infinite-volume fit results in HMrS$\chi$PT.

\section{Outlook}\label{sec:outro}

In this work, we have shown preliminary results for the blinded form factors
$f_+$, $f_0$, and $f_T$ in the chiral continuum, using $B_s \to K$ form factors
to illustrate our findings.
Results for the other decays, $B \to \pi$ and $B \to K$, are similar both in
general shape and in systematic effects.
We are extrapolating in a model-independent manner to the full kinematic range
accessible in experiment by using the
functional implementation \cite{Lattice:2015tia} of the
BCL parametrization \cite{Bourrely:2008za} of the
$z$ expansion \cite{Boyd:1994tt}.
Such $z$-expansion studies are ongoing.
This work serves as a successor to the earlier asqtad analyses
\cite{Bazavov:2019aom, Bailey:2015dka, Bailey:2008wp} and aims to reduce the
errors due to scale setting and chiral-continuum fitting.

\section*{Acknowledgments}\label{sec:thanks}

This work was supported in part
by the U.S.~Department of Energy,
by the U.S.~National Science Foundation,
by the Fermilab Distinguished Scholars Program (A.X.K.),
by German Excellence Initiative and the European Union Seventh Framework Program
    as well as the European Union's Marie Curie COFUND program (A.S.K.),
by the Blue Waters PAID program (Y.L.), and
by the Visiting Scholars Program of the Universities Research Association
    (Z.G., Y.L.).
Computations for this work were carried out with resources provided
by the USQCD Collaboration;
by the ALCF and NERSC, which are funded by the U.S.~Department of Energy; and
by NCAR, NCSA, NICS, TACC, and Blue Waters, which are funded through the
    U.S.~National Science Foundation.
This research is part of the Blue Waters sustained-petascale computing project,
which is supported by the National Science Foundation (awards OCI-0725070 and
ACI-1238993) and the state of Illinois. Blue Waters is a joint effort of the
University of Illinois at Urbana--Champaign and its National Center for
Supercomputing Applications.
Fermilab is operated by Fermi Research Alliance, LLC under Contract
No.~DE-AC02-07CH11359 with the United States Department of Energy, Office of
Science, Office of High Energy Physics.

\bibliographystyle{JHEP}
\bibliography{lattice2019}

\end{document}